\documentclass[
reprint,
superscriptaddress,
preprintnumbers,
amsmath,
amssymb,
aps,
prd,
twocolumn
]{revtex4-2}

\usepackage{graphicx}
\usepackage{dcolumn}
\usepackage{bm}
\usepackage[utf8]{inputenc}
\usepackage{tabularx}
\usepackage[dvipsnames]{xcolor}
\usepackage[euler]{textgreek}
\usepackage{upgreek}
\usepackage{derivative}
\usepackage{amsmath}
\usepackage{physics}
\usepackage{amssymb}
\usepackage{hyperref}
\hypersetup{colorlinks=true,linkcolor=blue!50!black,urlcolor=blue!50!black,citecolor=blue!50!black}
\usepackage{colortbl}
\usepackage{svg}
\usepackage{romannum}
\usepackage{booktabs}
\usepackage{array}
\usepackage{makecell}
\usepackage[margin=1in]{geometry}  
\usepackage{dsfont}
\usepackage[caption=false]{subfig}
\usepackage{siunitx}

\newcommand{\OPER}{\mathcal{O}}

\newcommand{\Tx}{T_\mathrm{\chi}}
\newcommand{\Ti}{T_i}
\newcommand{\mx}{m_\mathrm{\chi}}
\newcommand{\mi}{m_i}
\newcommand{\Er}{E_\mathrm{R}}

\newcommand\freefootnote[1]{%
  \let\thefootnote\relax%
  \footnotetext{#1}%
  \let\thefootnote\svthefootnote%
}

\begin{document}

\title{Cosmic ray boosted dark matter with momentum dependent interactions can explain the LZ 248 keV event}

\author{Matti Heikinheimo}
\email{matti.heikinheimo@helsinki.fi}
\affiliation{Department of Physics, P.O.Box 64, FI-00014 University of Helsinki, Finland}
\affiliation{Helsinki Institute of Physics, P.O.Box 64, FI-00014 University of Helsinki, Finland}

\author{Niklas Zimmermann}
\email{niklas.zimmermann@helsinki.fi}
\affiliation{Department of Physics, P.O.Box 64, FI-00014 University of Helsinki, Finland}
\affiliation{Helsinki Institute of Physics, P.O.Box 64, FI-00014 University of Helsinki, Finland}


\begin{abstract}
The LZ experiment has recently reported an observation of a single event with characteristics of a nuclear recoil at a high recoil energy of 248 keV, where no background events are expected. We show that such nuclear recoil event could be caused by high velocity dark matter particles originating from interactions with energetic cosmic rays. We analyze the expected recoil spectrum due to cosmic ray boosted dark matter within the non-relativistic effective theory formalism, and show that momentum-dependent operators, such as $\mathcal{O}_6$ and $\mathcal{O}_{10}$, can produce a spectrum that peaks at the observed energy.
\end{abstract}

\maketitle
\section{Introduction}

The LUX-ZEPLIN (LZ) experiment has observed a single event in a high recoil energy search \cite{LZ:2026axp} that matches the characteristics of a nuclear recoil with 248 $\pm$ 23 (stat) $\pm$ 23 (sys) keV recoil energy. In typical WIMP models, where the dark matter (DM) particles originate directly from the DM halo, such high energy events would be expected to be accompanied by a larger number of events at lower energies. However, no nuclear recoil events in excess of the expected background are observed at lower energy.

A recoil spectrum peaking at high recoil energy could be caused by inelastic DM scattering, as discussed e.g. in \cite{Fan:2026kxx,Freese:2026sga,deLima:2026shq,Su:2026rwz,McCabe:2026crm,Wu:2026nhi,Baer:2026fpy,DiMauro:2026ldr,Visinelli:2026kgt,Yin:2026jnn,Dent:2026bji,Smirnov:2026aqk,Chattopadhyay:2026ryw,Du:2026guj,Rodd:2026tyn,Gu:2026vto,Wang:2026ytg,Yamashita:2026ump,Nomura:2026qyq,asadi2026xenonrecoilsmagneticinelastic}. However, some of these models are subject to constraints from solar capture, which would result in neutrino emission due to annihilation of the captured DM particles \cite{Pospelov:2026ewn,DiMauro:2026dqp,Bose:2026ndd}. Another possibility is via momentum dependent interactions, which can result in a more complex spectrum compared to the spin-independent coherent scattering \cite{DiMauro:2026ldr}, as was discussed also in the analysis by the LZ collaboration \cite{LZ:2026axp}. However, the resulting spectra in these scenarios still contain a significant component below the observed energy, so that the non-observation of any events in the low energy region is difficult to explain.

Boosted DM is a scenario, where a smaller population of DM particles exhibit velocities in excess of the galactic escape velocity. Boosted DM originating from decays of a heavier parent particle have been explored as a possible explanation for the LZ event in \cite{Alhazmi:2026efz,Liang:2026coz,Kannike:2026qyl}. In this work we analyze a scenario, where the boosted DM originates from scattering with energetic cosmic rays. We emphasize that this is a very minimal framework, as the cosmic ray boosted component arises from the same DM-nucleon interactions that give rise to the nuclear recoil signal in the detectors. Therefore the framework requires no additional assumptions about the spectrum or interactions within the dark sector, contrary to the scenarios of inelastic scattering or boosted DM due to decays of a parent particle. Our analysis is based on the non-relativistic effective theory of DM-nucleus interactions \cite{Fitzpatrick:2012ix}, where we focus on the momentum-dependent operators which naturally give rise to a spectrum peaking at high recoil energy.

\section{Cosmic ray boosted DM in non-relativistic effective theory}

Cosmic ray boosted DM was first discussed in \cite{Bringmann:2018cvk} in context of light dark matter, where the motivation is to boost part of the recoil spectrum above the detection threshold which otherwise limits the sensitivity of the direct detection experiments to light DM. This formalism has later been developed to cover various energy-dependent DM-nucleus cross sections, using a classification based on simplified models in \cite{COSINUS:2026acs}. In this work we follow the formalism used in \cite{COSINUS:2026acs}, but instead of relativistic simplified Lagrangians, we work in the non-relativistic limit described by the nuclear response function formalism of \cite{Fitzpatrick:2012ix,Anand:2013yka}. This is justified, as we consider the DM mass in the GeV range, which ensures that even after the boost the DM particles remain non-relativistic or at most weakly relativistic, as can be seen in the spectra reported in~\cite{COSINUS:2026acs}. Furthermore, where the momentum-dependent operators mostly affect the results is in the form factors of the Xenon nuclei, and the $\mathcal{O}$(100) KeV nuclear recoil energies are well within the region of validity for the effective theory for the heavy Xenon nucleus. In fact, even if the DM particle is lighter and thus relativistic, we expect the effective theory framework to remain mostly valid as a description of the nuclear response. We note that the effective operators $\OPER_6$ and $\OPER_{10}$ arise naturally from the Lorentz invariant operators $\bar{\chi}\gamma^5\chi\bar{n}\gamma^5 n$ and $\bar{\chi}\chi\bar{n}\gamma^5 n$, respectively, where $\chi$ is the DM spinor and $n$ the nucleon spinor \cite{DelNobile:2018dfg}.

The boosted DM flux is given by \cite{COSINUS:2026acs}
\begin{equation}
     \frac{d\Phi_\mathrm{\chi}}{d \Tx} = D_\mathrm{eff} \frac{\rho_\chi}{\mx} \int_{\Ti^{\mathrm{min}}}^\infty \frac{d\Phi_\mathrm{\mathrm{LIS}}}{d \Ti} \frac{d \sigma_{\chi i}}{d \Tx} d\Ti,
\end{equation}
where $D_\mathrm{eff} = \SI{8.02}{kpc}$ \cite{Bringmann:2018cvk} is the effective distance, $\rho_\chi = \SI{0.4}{GeV/cm^3}$ \cite{Cirelli:2024ssz} the local DM density and $d\Phi_\mathrm{LIS}/d\Ti$ is the differential cosmic ray flux, given in \cite{DellaTorre:2016jjf,Boschini:2017fxq}. The differential DM-nucleon cross section can be written in the non-relativistic effective theory as \cite{Anand:2013yka}
\begin{equation}
    \frac{d \sigma_{\chi i}}{d \Tx} = \frac{\mx\mi}{4\pi\Ti}\left[ \frac{1}{2j_i+1}\frac{1}{2j_\chi+1}\sum_{\rm spin}|\mathcal{M}^{\rm Nuc}|^2 \right],
    \label{eq:diffxsec}
\end{equation}
where $j_i$ is the spin of the nucleus, $j_\chi$ the spin of the DM particle and $\mathcal{M}^{\rm Nuc}$ is the DM-nucleus scattering amplitude given in terms of the nuclear response functions in \cite{Anand:2013yka} for Xenon and in \cite{Catena:2015uha} for the cosmic ray nuclei H and He. For the operators $\OPER_6$ and $\OPER_{10}$ we consider, only the proton flux is relevant, since the coupling to He nuclei vanish for these operators.

The nuclear recoil event rate in a DM detector is obtained in terms of the boosted DM flux as \cite{COSINUS:2026acs}
\begin{equation}
    \frac{dR}{d E_\mathrm{R}} = \sum_{t} \epsilon N_t  \int_{\Tx^{\mathrm{min}}}^\infty \frac{d\Phi_\mathrm{\chi}}{d \Tx} \frac{d \sigma_{\chi t}}{d \Er} d\Tx,
\end{equation}
where the sum is over the target isotopes $t$, $\epsilon$ is the detection efficiency, $N_t$ is the number of target isotopes $t$ per kg and the differential cross section $d\sigma_{\chi t}/d\Er$ is obtained from equation (\ref{eq:diffxsec}) via the replacement $\Tx\rightarrow \Er$, $\Ti\rightarrow\Tx$, $m_\chi\rightarrow m_t$, $m_i\rightarrow m_\chi$, $j_i\rightarrow j_t$.

\begin{figure}[t]
    \centering
    \includegraphics[width = \linewidth]{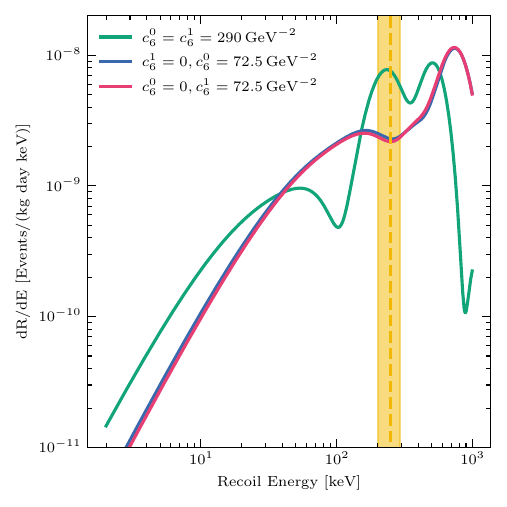}
    \caption{Differential nuclear recoil event rate in Xenon for cosmic ray boosted DM due to $\OPER_6$ interactions for $m_\chi=1$ GeV. The recoil energy matching the event observed by LZ is marked by the yellow dashed line. The unertainties are portraid by the yellow shaded region.}
    \label{fig:spectrumO6}
\end{figure}

\begin{figure}[h]
    \centering
    \includegraphics[width = \linewidth]{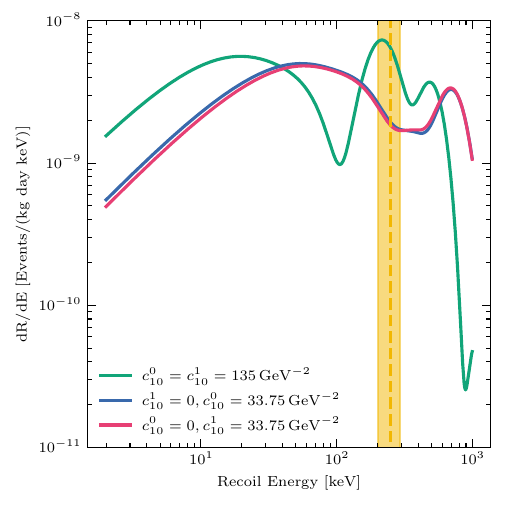}
    \caption{Differential nuclear recoil event rate in Xenon for cosmic ray boosted DM due to $\OPER_{10}$ interactions for $m_\chi=1$ GeV. The recoil energy matching the event observed by LZ is marked by the yellow dashed line. The unertainties are portraid by the yellow shaded region.}
    \label{fig:spectrumO10}
\end{figure}

The resulting nuclear recoil event rate in a Xenon target for the operator $\OPER_6$ is shown in figure \ref{fig:spectrumO6}, for $m_\chi=1$ GeV, $c_6^0=c_6^1 = 290\,\mathrm{GeV}^{-2}$. This choice of Wilson coefficients results in one expected event within the window from $100\,\mathrm{keV}$ to $300\,\mathrm{keV}$ using the 2.84 ton year exposure of the LZ extended recoil energy search. We observe that a local maximum of the spectrum almost precisely coincides with the observed event, while the rate at lower energy is suppressed by more that one order of magnitude. However, there is a second maximum of similar magnitude at even higher energy around 500 keV. For comparison, we also show the spectra for $c_6^0=0$ and $c_6^1=0$ where the non-zero coefficient is set to $72.5\,\mathrm{GeV}^{-2}$. We observe that the fit is clearly worse than the $c_6^0=c_6^1$ case, as the maximum of the spectrum is now well above the region of interest.

The spectrum for the operator $\mathcal{O}_{10}$ is shown in figure \ref{fig:spectrumO10}, where we also observe a maximum close to the observed energy. For this scenario, the suppression of the event rate in the low energy region is weaker compared to $\mathcal{O}_6$. This can be quantified as follows: Using the normalization that yields an expectation value of one event within the 100 keV to 300 keV window, operator $\mathcal{O}_6$ yields an expectation of $0.069$ events below $100\,\mathrm{keV}$. For $\OPER_{10}$ the this number is $0.35$ events, but correspondingly the spectrum is more suppressed in the region above 300 keV. More events would be required to make further inferences about the shape of the spectrum and thus on the relative values of the coupling coefficients. The cosmic ray boosted spectrum is not very sensitive to the DM mass, and therefore not much can be said about the DM mass based on the single event. However, to avoid constraints due to the non-boosted halo component, the DM mass should be below a few GeV.

\section{Conclusions}
We have shown that cosmic ray boosted DM with momentum dependent interactions can naturally produce a recoil spectrum in Xenon which peaks close to the high energy nuclear recoil event observed in the LZ experiment. The operators $\OPER_6$ and $\OPER_{10}$ seem to give the best fits, but obviously more events are required to reconstruct the spectrum and thereby constrain the model better. If the excess persists in future data and in other Xenon-based experiments, the non-relativistic effective theory framework can be used to find the models that best describe the observed recoil spectrum. The cosmic ray boosted DM with momentum dependent interactions is a natural scenario that can explain a hard recoil spectrum without additional assumptions about the mass spectrum of the dark sector.

\section*{Acknowledgments}
This work has been supported by the Research Council of Finland (grant$\#$ 371542) and by the Finnish Graduate School in Particle and Nuclear Physics (Doctoral Education Pilot).

\bibliography{Bibliography.bib}

\begin{thebibliography}{36}%
\makeatletter
\providecommand \@ifxundefined [1]{%
 \@ifx{#1\undefined}
}%
\providecommand \@ifnum [1]{%
 \ifnum #1\expandafter \@firstoftwo
 \else \expandafter \@secondoftwo
 \fi
}%
\providecommand \@ifx [1]{%
 \ifx #1\expandafter \@firstoftwo
 \else \expandafter \@secondoftwo
 \fi
}%
\providecommand \natexlab [1]{#1}%
\providecommand \enquote  [1]{``#1''}%
\providecommand \bibnamefont  [1]{#1}%
\providecommand \bibfnamefont [1]{#1}%
\providecommand \citenamefont [1]{#1}%
\providecommand \href@noop [0]{\@secondoftwo}%
\providecommand \href [0]{\begingroup \@sanitize@url \@href}%
\providecommand \@href[1]{\@@startlink{#1}\@@href}%
\providecommand \@@href[1]{\endgroup#1\@@endlink}%
\providecommand \@sanitize@url [0]{\catcode `\\12\catcode `\$12\catcode
  `\&12\catcode `\#12\catcode `\^12\catcode `\_12\catcode `\%12\relax}%
\providecommand \@@startlink[1]{}%
\providecommand \@@endlink[0]{}%
\providecommand \url  [0]{\begingroup\@sanitize@url \@url }%
\providecommand \@url [1]{\endgroup\@href {#1}{\urlprefix }}%
\providecommand \urlprefix  [0]{URL }%
\providecommand \Eprint [0]{\href }%
\providecommand \doibase [0]{https://doi.org/}%
\providecommand \selectlanguage [0]{\@gobble}%
\providecommand \bibinfo  [0]{\@secondoftwo}%
\providecommand \bibfield  [0]{\@secondoftwo}%
\providecommand \translation [1]{[#1]}%
\providecommand \BibitemOpen [0]{}%
\providecommand \bibitemStop [0]{}%
\providecommand \bibitemNoStop [0]{.\EOS\space}%
\providecommand \EOS [0]{\spacefactor3000\relax}%
\providecommand \BibitemShut  [1]{\csname bibitem#1\endcsname}%
\let\auto@bib@innerbib\@empty
\bibitem [{\citenamefont {Akerib}\ \emph {et~al.}(2026)\citenamefont {Akerib}
  \emph {et~al.}}]{LZ:2026axp}%
  \BibitemOpen
  \bibfield  {author} {\bibinfo {author} {\bibfnamefont {D.~S.}\ \bibnamefont
  {Akerib}} \emph {et~al.} (\bibinfo {collaboration} {LZ}),\ }\bibfield
  {title} {\bibinfo {title} {{Search for dark matter particle interactions in
  an extended nuclear recoil energy window with the LUX-ZEPLIN (LZ)
  experiment}},\ }\href@noop {} {\  (\bibinfo {year} {2026})},\ \Eprint
  {https://arxiv.org/abs/2609.02823} {arXiv:2609.02823 [hep-ex]} \BibitemShut
  {NoStop}%
\bibitem [{\citenamefont {Fan}\ and\ \citenamefont
  {Reece}(2026)}]{Fan:2026kxx}%
  \BibitemOpen
  \bibfield  {author} {\bibinfo {author} {\bibfnamefont {J.}~\bibnamefont
  {Fan}}\ and\ \bibinfo {author} {\bibfnamefont {M.}~\bibnamefont {Reece}},\
  }\bibfield  {title} {\bibinfo {title} {{Higgsino Above the Sea of Fog}},\
  }\href@noop {} {\  (\bibinfo {year} {2026})},\ \Eprint
  {https://arxiv.org/abs/2609.01504} {arXiv:2609.01504 [hep-ph]} \BibitemShut
  {NoStop}%
\bibitem [{\citenamefont {Freese}\ and\ \citenamefont
  {Theodosopoulos}(2026)}]{Freese:2026sga}%
  \BibitemOpen
  \bibfield  {author} {\bibinfo {author} {\bibfnamefont {K.}~\bibnamefont
  {Freese}}\ and\ \bibinfo {author} {\bibfnamefont {D.~P.}\ \bibnamefont
  {Theodosopoulos}},\ }\bibfield  {title} {\bibinfo {title} {{Higgsino Dark
  Matter Interpretation of the LUX-ZEPLIN 248 keV Nuclear-Recoil Event}},\
  }\href@noop {} {\  (\bibinfo {year} {2026})},\ \Eprint
  {https://arxiv.org/abs/2609.01583} {arXiv:2609.01583 [hep-ph]} \BibitemShut
  {NoStop}%
\bibitem [{\citenamefont {de~Lima}(2026)}]{deLima:2026shq}%
  \BibitemOpen
  \bibfield  {author} {\bibinfo {author} {\bibfnamefont {C.~H.}\ \bibnamefont
  {de~Lima}},\ }\bibfield  {title} {\bibinfo {title} {{Exothermic Dark Matter
  at LZ}},\ }\href@noop {} {\  (\bibinfo {year} {2026})},\ \Eprint
  {https://arxiv.org/abs/2609.05204} {arXiv:2609.05204 [hep-ph]} \BibitemShut
  {NoStop}%
\bibitem [{\citenamefont {Su}\ \emph {et~al.}(2026)\citenamefont {Su},
  \citenamefont {Yang},\ and\ \citenamefont {Yang}}]{Su:2026rwz}%
  \BibitemOpen
  \bibfield  {author} {\bibinfo {author} {\bibfnamefont {L.}~\bibnamefont
  {Su}}, \bibinfo {author} {\bibfnamefont {J.~M.}\ \bibnamefont {Yang}},\ and\
  \bibinfo {author} {\bibfnamefont {W.-N.}\ \bibnamefont {Yang}},\ }\bibfield
  {title} {\bibinfo {title} {{Inelastic Dark Matter Signature at High Recoil
  Energy in LUX-ZEPLIN and CRESST}},\ }\href@noop {} {\  (\bibinfo {year}
  {2026})},\ \Eprint {https://arxiv.org/abs/2609.01475} {arXiv:2609.01475
  [hep-ph]} \BibitemShut {NoStop}%
\bibitem [{\citenamefont {McCabe}(2026)}]{McCabe:2026crm}%
  \BibitemOpen
  \bibfield  {author} {\bibinfo {author} {\bibfnamefont {C.}~\bibnamefont
  {McCabe}},\ }\bibfield  {title} {\bibinfo {title} {{Seasonal dark matter from
  the LUX-ZEPLIN high-energy event}},\ }\href@noop {} {\  (\bibinfo {year}
  {2026})},\ \Eprint {https://arxiv.org/abs/2609.04181} {arXiv:2609.04181
  [hep-ph]} \BibitemShut {NoStop}%
\bibitem [{\citenamefont {Wu}\ \emph {et~al.}(2026)\citenamefont {Wu},
  \citenamefont {Zhang},\ and\ \citenamefont {Zhu}}]{Wu:2026nhi}%
  \BibitemOpen
  \bibfield  {author} {\bibinfo {author} {\bibfnamefont {L.}~\bibnamefont
  {Wu}}, \bibinfo {author} {\bibfnamefont {Y.}~\bibnamefont {Zhang}},\ and\
  \bibinfo {author} {\bibfnamefont {B.}~\bibnamefont {Zhu}},\ }\bibfield
  {title} {\bibinfo {title} {{TeV Higgsino Dark Matter from LZ Nuclear Recoil
  to Fermi-LAT Gamma Rays}},\ }\href@noop {} {\  (\bibinfo {year} {2026})},\
  \Eprint {https://arxiv.org/abs/2609.01590} {arXiv:2609.01590 [hep-ph]}
  \BibitemShut {NoStop}%
\bibitem [{\citenamefont {Baer}\ and\ \citenamefont
  {Barger}(2026)}]{Baer:2026fpy}%
  \BibitemOpen
  \bibfield  {author} {\bibinfo {author} {\bibfnamefont {H.}~\bibnamefont
  {Baer}}\ and\ \bibinfo {author} {\bibfnamefont {V.}~\bibnamefont {Barger}},\
  }\bibfield  {title} {\bibinfo {title} {{Exothermic dark matter and the 248
  keV nuclear recoil in LUX-ZEPLIN}},\ }\href@noop {} {\  (\bibinfo {year}
  {2026})},\ \Eprint {https://arxiv.org/abs/2609.06153} {arXiv:2609.06153
  [hep-ph]} \BibitemShut {NoStop}%
\bibitem [{\citenamefont {Di~Mauro}(2026)}]{DiMauro:2026ldr}%
  \BibitemOpen
  \bibfield  {author} {\bibinfo {author} {\bibfnamefont {M.}~\bibnamefont
  {Di~Mauro}},\ }\bibfield  {title} {\bibinfo {title} {{Dark Matter at the
  Kinematic Edge: Interpreting the 248 keV LZ Nuclear-Recoil Candidate}},\
  }\href@noop {} {\  (\bibinfo {year} {2026})},\ \Eprint
  {https://arxiv.org/abs/2609.02608} {arXiv:2609.02608 [hep-ph]} \BibitemShut
  {NoStop}%
\bibitem [{\citenamefont {Visinelli}(2026)}]{Visinelli:2026kgt}%
  \BibitemOpen
  \bibfield  {author} {\bibinfo {author} {\bibfnamefont {L.}~\bibnamefont
  {Visinelli}},\ }\bibfield  {title} {\bibinfo {title} {{A Peccei-Quinn Origin
  for Inelastic Electroweak Dark Matter after LUX-ZEPLIN}},\ }\href@noop {} {\
  (\bibinfo {year} {2026})},\ \Eprint {https://arxiv.org/abs/2609.02807}
  {arXiv:2609.02807 [hep-ph]} \BibitemShut {NoStop}%
\bibitem [{\citenamefont {Yin}(2026)}]{Yin:2026jnn}%
  \BibitemOpen
  \bibfield  {author} {\bibinfo {author} {\bibfnamefont {W.}~\bibnamefont
  {Yin}},\ }\bibfield  {title} {\bibinfo {title} {{A PQ-Symmetric High-Scale
  SUSY Interpretation of the LZ High-Energy Recoil}},\ }\href@noop {} {\
  (\bibinfo {year} {2026})},\ \Eprint {https://arxiv.org/abs/2609.01892}
  {arXiv:2609.01892 [hep-ph]} \BibitemShut {NoStop}%
\bibitem [{\citenamefont {Dent}\ and\ \citenamefont
  {Newstead}(2026)}]{Dent:2026bji}%
  \BibitemOpen
  \bibfield  {author} {\bibinfo {author} {\bibfnamefont {J.~B.}\ \bibnamefont
  {Dent}}\ and\ \bibinfo {author} {\bibfnamefont {J.~L.}\ \bibnamefont
  {Newstead}},\ }\bibfield  {title} {\bibinfo {title} {{Exothermic and
  Endothermic Inelastic Dark Matter Interpretations at LZ: Sideband Constraints
  and Future Prospects}},\ }\href@noop {} {\  (\bibinfo {year} {2026})},\
  \Eprint {https://arxiv.org/abs/2609.04673} {arXiv:2609.04673 [hep-ph]}
  \BibitemShut {NoStop}%
\bibitem [{\citenamefont {Smirnov}\ \emph {et~al.}(2026)\citenamefont
  {Smirnov}, \citenamefont {Griffith},\ and\ \citenamefont
  {Beacom}}]{Smirnov:2026aqk}%
  \BibitemOpen
  \bibfield  {author} {\bibinfo {author} {\bibfnamefont {J.}~\bibnamefont
  {Smirnov}}, \bibinfo {author} {\bibfnamefont {S.}~\bibnamefont {Griffith}},\
  and\ \bibinfo {author} {\bibfnamefont {J.~F.}\ \bibnamefont {Beacom}},\
  }\bibfield  {title} {\bibinfo {title} {{Inelastic Signatures of Electroweak
  Dark Matter}},\ }\href@noop {} {\  (\bibinfo {year} {2026})},\ \Eprint
  {https://arxiv.org/abs/2609.04144} {arXiv:2609.04144 [hep-ph]} \BibitemShut
  {NoStop}%
\bibitem [{\citenamefont {Chattopadhyay}\ \emph {et~al.}(2026)\citenamefont
  {Chattopadhyay}, \citenamefont {Das}, \citenamefont {Puri},\ and\
  \citenamefont {Roy}}]{Chattopadhyay:2026ryw}%
  \BibitemOpen
  \bibfield  {author} {\bibinfo {author} {\bibfnamefont {U.}~\bibnamefont
  {Chattopadhyay}}, \bibinfo {author} {\bibfnamefont {D.}~\bibnamefont {Das}},
  \bibinfo {author} {\bibfnamefont {R.}~\bibnamefont {Puri}},\ and\ \bibinfo
  {author} {\bibfnamefont {J.}~\bibnamefont {Roy}},\ }\bibfield  {title}
  {\bibinfo {title} {{Sub-TeV Singlino Dark Matter in light from Sagittarius
  A$^\ast$ and LUX-ZEPLIN Nuclear-Recoil Event}},\ }\href@noop {} {\  (\bibinfo
  {year} {2026})},\ \Eprint {https://arxiv.org/abs/2609.02994}
  {arXiv:2609.02994 [hep-ph]} \BibitemShut {NoStop}%
\bibitem [{\citenamefont {Du}\ and\ \citenamefont {Wang}(2026)}]{Du:2026guj}%
  \BibitemOpen
  \bibfield  {author} {\bibinfo {author} {\bibfnamefont {X.}~\bibnamefont
  {Du}}\ and\ \bibinfo {author} {\bibfnamefont {F.}~\bibnamefont {Wang}},\
  }\bibfield  {title} {\bibinfo {title} {{TeV Higgsino Interpretation of the LZ
  High-Recoil Event with Intermediate-Scale Electroweak Gauginos}},\
  }\href@noop {} {\  (\bibinfo {year} {2026})},\ \Eprint
  {https://arxiv.org/abs/2609.04163} {arXiv:2609.04163 [hep-ph]} \BibitemShut
  {NoStop}%
\bibitem [{\citenamefont {Rodd}\ \emph {et~al.}(2026)\citenamefont {Rodd},
  \citenamefont {Safdi}, \citenamefont {Slatyer},\ and\ \citenamefont
  {Xu}}]{Rodd:2026tyn}%
  \BibitemOpen
  \bibfield  {author} {\bibinfo {author} {\bibfnamefont {N.~L.}\ \bibnamefont
  {Rodd}}, \bibinfo {author} {\bibfnamefont {B.~R.}\ \bibnamefont {Safdi}},
  \bibinfo {author} {\bibfnamefont {T.~R.}\ \bibnamefont {Slatyer}},\ and\
  \bibinfo {author} {\bibfnamefont {W.~L.}\ \bibnamefont {Xu}},\ }\bibfield
  {title} {\bibinfo {title} {{Confronting the Higgsino Interpretation of the LZ
  Event with the High-Energy Sideband}},\ }\href@noop {} {\  (\bibinfo {year}
  {2026})},\ \Eprint {https://arxiv.org/abs/2609.04175} {arXiv:2609.04175
  [hep-ph]} \BibitemShut {NoStop}%
\bibitem [{\citenamefont {Gu}\ \emph {et~al.}(2026)\citenamefont {Gu},
  \citenamefont {Li}, \citenamefont {Tang},\ and\ \citenamefont
  {Xu}}]{Gu:2026vto}%
  \BibitemOpen
  \bibfield  {author} {\bibinfo {author} {\bibfnamefont {G.}~\bibnamefont
  {Gu}}, \bibinfo {author} {\bibfnamefont {L.}~\bibnamefont {Li}}, \bibinfo
  {author} {\bibfnamefont {S.-S.}\ \bibnamefont {Tang}},\ and\ \bibinfo
  {author} {\bibfnamefont {Y.}~\bibnamefont {Xu}},\ }\bibfield  {title}
  {\bibinfo {title} {{Inelastic from the Other Side: Xenon Excitation Signals
  in Light of the LZ High-Recoil Event}},\ }\href@noop {} {\  (\bibinfo {year}
  {2026})},\ \Eprint {https://arxiv.org/abs/2609.05291} {arXiv:2609.05291
  [hep-ph]} \BibitemShut {NoStop}%
\bibitem [{\citenamefont {Wang}\ and\ \citenamefont
  {Xiao}(2026)}]{Wang:2026ytg}%
  \BibitemOpen
  \bibfield  {author} {\bibinfo {author} {\bibfnamefont {L.}~\bibnamefont
  {Wang}}\ and\ \bibinfo {author} {\bibfnamefont {Y.}~\bibnamefont {Xiao}},\
  }\bibfield  {title} {\bibinfo {title} {{The Inert Doublet Model of Dark
  Matter and the LUX-ZEPLIN High-Recoil Event}},\ }\href@noop {} {\  (\bibinfo
  {year} {2026})},\ \Eprint {https://arxiv.org/abs/2609.06571}
  {arXiv:2609.06571 [hep-ph]} \BibitemShut {NoStop}%
\bibitem [{\citenamefont {Yamashita}(2026)}]{Yamashita:2026ump}%
  \BibitemOpen
  \bibfield  {author} {\bibinfo {author} {\bibfnamefont {K.}~\bibnamefont
  {Yamashita}},\ }\bibfield  {title} {\bibinfo {title} {{Inelastic Dark Photon
  Dark Matter for the LUX-ZEPLIN High-Recoil Event and the Galactic Halo
  Gamma-Ray Excess}},\ }\href@noop {} {\  (\bibinfo {year} {2026})},\ \Eprint
  {https://arxiv.org/abs/2609.02868} {arXiv:2609.02868 [hep-ph]} \BibitemShut
  {NoStop}%
\bibitem [{\citenamefont {Nomura}(2026)}]{Nomura:2026qyq}%
  \BibitemOpen
  \bibfield  {author} {\bibinfo {author} {\bibfnamefont {Y.}~\bibnamefont
  {Nomura}},\ }\bibfield  {title} {\bibinfo {title} {{Dark Matter as the Z{\_}2
  Partner of the Standard Model Higgs Boson}},\ }\href@noop {} {\  (\bibinfo
  {year} {2026})},\ \Eprint {https://arxiv.org/abs/2609.02505}
  {arXiv:2609.02505 [hep-ph]} \BibitemShut {NoStop}%
\bibitem [{\citenamefont {Asadi}\ \emph {et~al.}(2026)\citenamefont {Asadi},
  \citenamefont {Batz}, \citenamefont {Fox}, \citenamefont {Homiller},\ and\
  \citenamefont {Kribs}}]{asadi2026xenonrecoilsmagneticinelastic}%
  \BibitemOpen
  \bibfield  {author} {\bibinfo {author} {\bibfnamefont {P.}~\bibnamefont
  {Asadi}}, \bibinfo {author} {\bibfnamefont {A.}~\bibnamefont {Batz}},
  \bibinfo {author} {\bibfnamefont {P.~J.}\ \bibnamefont {Fox}}, \bibinfo
  {author} {\bibfnamefont {S.~D.}\ \bibnamefont {Homiller}},\ and\ \bibinfo
  {author} {\bibfnamefont {G.~D.}\ \bibnamefont {Kribs}},\ }\href
  {https://arxiv.org/abs/2609.09107} {\bibinfo {title} {For whom the xenon
  recoils: Magnetic inelastic dark baryons}} (\bibinfo {year} {2026}),\ \Eprint
  {https://arxiv.org/abs/2609.09107} {arXiv:2609.09107 [hep-ph]} \BibitemShut
  {NoStop}%
\bibitem [{\citenamefont {Pospelov}\ and\ \citenamefont
  {Ramani}(2026)}]{Pospelov:2026ewn}%
  \BibitemOpen
  \bibfield  {author} {\bibinfo {author} {\bibfnamefont {M.}~\bibnamefont
  {Pospelov}}\ and\ \bibinfo {author} {\bibfnamefont {H.}~\bibnamefont
  {Ramani}},\ }\bibfield  {title} {\bibinfo {title} {{Strong Constraints on
  Higgsino Dark Matter from Solar Capture}},\ }\href@noop {} {\  (\bibinfo
  {year} {2026})},\ \Eprint {https://arxiv.org/abs/2609.02775}
  {arXiv:2609.02775 [hep-ph]} \BibitemShut {NoStop}%
\bibitem [{\citenamefont {Di~Mauro}\ and\ \citenamefont
  {Shaikh}(2026)}]{DiMauro:2026dqp}%
  \BibitemOpen
  \bibfield  {author} {\bibinfo {author} {\bibfnamefont {M.}~\bibnamefont
  {Di~Mauro}}\ and\ \bibinfo {author} {\bibfnamefont {H.}~\bibnamefont
  {Shaikh}},\ }\bibfield  {title} {\bibinfo {title} {{Solar Capture Tests of
  Inelastic Dark Matter after the LZ High-Recoil Event}},\ }\href@noop {} {\
  (\bibinfo {year} {2026})},\ \Eprint {https://arxiv.org/abs/2609.06760}
  {arXiv:2609.06760 [hep-ph]} \BibitemShut {NoStop}%
\bibitem [{\citenamefont {Bose}\ \emph {et~al.}(2026)\citenamefont {Bose} \emph
  {et~al.}}]{Bose:2026ndd}%
  \BibitemOpen
  \bibfield  {author} {\bibinfo {author} {\bibfnamefont {D.}~\bibnamefont
  {Bose}} \emph {et~al.},\ }\bibfield  {title} {\bibinfo {title} {{Not so good
  $ν$s for Higgsino dark matter as LZ excess: stringent limits from
  Super-Kamiokande and IceCube}},\ }\href@noop {} {\  (\bibinfo {year}
  {2026})},\ \Eprint {https://arxiv.org/abs/2609.07807} {arXiv:2609.07807
  [hep-ph]} \BibitemShut {NoStop}%
\bibitem [{\citenamefont {Alhazmi}\ \emph {et~al.}(2026)\citenamefont
  {Alhazmi}, \citenamefont {Kim}, \citenamefont {Kong}, \citenamefont {Park},\
  and\ \citenamefont {Shin}}]{Alhazmi:2026efz}%
  \BibitemOpen
  \bibfield  {author} {\bibinfo {author} {\bibfnamefont {H.}~\bibnamefont
  {Alhazmi}}, \bibinfo {author} {\bibfnamefont {D.}~\bibnamefont {Kim}},
  \bibinfo {author} {\bibfnamefont {K.}~\bibnamefont {Kong}}, \bibinfo {author}
  {\bibfnamefont {J.-C.}\ \bibnamefont {Park}},\ and\ \bibinfo {author}
  {\bibfnamefont {S.}~\bibnamefont {Shin}},\ }\bibfield  {title} {\bibinfo
  {title} {{High-Energy Nuclear Recoils from Boosted Dark Matter for the LZ
  248-keV Event: Beyond the Halo-Dependent High-Velocity Tail}},\ }\href@noop
  {} {\  (\bibinfo {year} {2026})},\ \Eprint {https://arxiv.org/abs/2609.06890}
  {arXiv:2609.06890 [hep-ph]} \BibitemShut {NoStop}%
\bibitem [{\citenamefont {Liang}\ \emph {et~al.}(2026)\citenamefont {Liang},
  \citenamefont {Liu}, \citenamefont {Tran},\ and\ \citenamefont
  {Xu}}]{Liang:2026coz}%
  \BibitemOpen
  \bibfield  {author} {\bibinfo {author} {\bibfnamefont {J.-H.}\ \bibnamefont
  {Liang}}, \bibinfo {author} {\bibfnamefont {Z.}~\bibnamefont {Liu}}, \bibinfo
  {author} {\bibfnamefont {V.~Q.}\ \bibnamefont {Tran}},\ and\ \bibinfo
  {author} {\bibfnamefont {Y.}~\bibnamefont {Xu}},\ }\bibfield  {title}
  {\bibinfo {title} {{LZ Nuclear-Recoil Excess from Boosted Light Magnetic
  Dipole-dipole Dark Matter}},\ }\href@noop {} {\  (\bibinfo {year} {2026})},\
  \Eprint {https://arxiv.org/abs/2609.06756} {arXiv:2609.06756 [hep-ph]}
  \BibitemShut {NoStop}%
\bibitem [{\citenamefont {Kannike}\ \emph {et~al.}(2026)\citenamefont
  {Kannike}, \citenamefont {Raidal},\ and\ \citenamefont
  {Strumia}}]{Kannike:2026qyl}%
  \BibitemOpen
  \bibfield  {author} {\bibinfo {author} {\bibfnamefont {K.}~\bibnamefont
  {Kannike}}, \bibinfo {author} {\bibfnamefont {M.}~\bibnamefont {Raidal}},\
  and\ \bibinfo {author} {\bibfnamefont {A.}~\bibnamefont {Strumia}},\
  }\bibfield  {title} {\bibinfo {title} {{Boosted dark particles and the LZ
  nuclear recoil event}},\ }\href@noop {} {\  (\bibinfo {year} {2026})},\
  \Eprint {https://arxiv.org/abs/2609.07742} {arXiv:2609.07742 [hep-ph]}
  \BibitemShut {NoStop}%
\bibitem [{\citenamefont {Fitzpatrick}\ \emph {et~al.}(2013)\citenamefont
  {Fitzpatrick}, \citenamefont {Haxton}, \citenamefont {Katz}, \citenamefont
  {Lubbers},\ and\ \citenamefont {Xu}}]{Fitzpatrick:2012ix}%
  \BibitemOpen
  \bibfield  {author} {\bibinfo {author} {\bibfnamefont {A.~L.}\ \bibnamefont
  {Fitzpatrick}}, \bibinfo {author} {\bibfnamefont {W.}~\bibnamefont {Haxton}},
  \bibinfo {author} {\bibfnamefont {E.}~\bibnamefont {Katz}}, \bibinfo {author}
  {\bibfnamefont {N.}~\bibnamefont {Lubbers}},\ and\ \bibinfo {author}
  {\bibfnamefont {Y.}~\bibnamefont {Xu}},\ }\bibfield  {title} {\bibinfo
  {title} {{The Effective Field Theory of Dark Matter Direct Detection}},\
  }\href {https://doi.org/10.1088/1475-7516/2013/02/004} {\bibfield  {journal}
  {\bibinfo  {journal} {JCAP}\ }\textbf {\bibinfo {volume} {02}},\ \bibinfo
  {pages} {004}},\ \Eprint {https://arxiv.org/abs/1203.3542} {arXiv:1203.3542
  [hep-ph]} \BibitemShut {NoStop}%
\bibitem [{\citenamefont {Bringmann}\ and\ \citenamefont
  {Pospelov}(2019)}]{Bringmann:2018cvk}%
  \BibitemOpen
  \bibfield  {author} {\bibinfo {author} {\bibfnamefont {T.}~\bibnamefont
  {Bringmann}}\ and\ \bibinfo {author} {\bibfnamefont {M.}~\bibnamefont
  {Pospelov}},\ }\bibfield  {title} {\bibinfo {title} {{Novel direct detection
  constraints on light dark matter}},\ }\href
  {https://doi.org/10.1103/PhysRevLett.122.171801} {\bibfield  {journal}
  {\bibinfo  {journal} {Phys. Rev. Lett.}\ }\textbf {\bibinfo {volume} {122}},\
  \bibinfo {pages} {171801} (\bibinfo {year} {2019})},\ \Eprint
  {https://arxiv.org/abs/1810.10543} {arXiv:1810.10543 [hep-ph]} \BibitemShut
  {NoStop}%
\bibitem [{\citenamefont {Angloher}\ \emph {et~al.}(2026)\citenamefont
  {Angloher} \emph {et~al.}}]{COSINUS:2026acs}%
  \BibitemOpen
  \bibfield  {author} {\bibinfo {author} {\bibfnamefont {G.}~\bibnamefont
  {Angloher}} \emph {et~al.} (\bibinfo {collaboration} {COSINUS}),\ }\bibfield
  {title} {\bibinfo {title} {{Cosmic Ray Boosted Dark Matter in COSINUS:
  Modeling and Constraints}},\ }\href@noop {} {\  (\bibinfo {year} {2026})},\
  \Eprint {https://arxiv.org/abs/2603.22971} {arXiv:2603.22971 [hep-ph]}
  \BibitemShut {NoStop}%
\bibitem [{\citenamefont {Anand}\ \emph {et~al.}(2014)\citenamefont {Anand},
  \citenamefont {Fitzpatrick},\ and\ \citenamefont {Haxton}}]{Anand:2013yka}%
  \BibitemOpen
  \bibfield  {author} {\bibinfo {author} {\bibfnamefont {N.}~\bibnamefont
  {Anand}}, \bibinfo {author} {\bibfnamefont {A.~L.}\ \bibnamefont
  {Fitzpatrick}},\ and\ \bibinfo {author} {\bibfnamefont {W.~C.}\ \bibnamefont
  {Haxton}},\ }\bibfield  {title} {\bibinfo {title} {{Weakly interacting
  massive particle-nucleus elastic scattering response}},\ }\href
  {https://doi.org/10.1103/PhysRevC.89.065501} {\bibfield  {journal} {\bibinfo
  {journal} {Phys. Rev. C}\ }\textbf {\bibinfo {volume} {89}},\ \bibinfo
  {pages} {065501} (\bibinfo {year} {2014})},\ \Eprint
  {https://arxiv.org/abs/1308.6288} {arXiv:1308.6288 [hep-ph]} \BibitemShut
  {NoStop}%
\bibitem [{\citenamefont {Del~Nobile}(2018)}]{DelNobile:2018dfg}%
  \BibitemOpen
  \bibfield  {author} {\bibinfo {author} {\bibfnamefont {E.}~\bibnamefont
  {Del~Nobile}},\ }\bibfield  {title} {\bibinfo {title} {{Complete
  Lorentz-to-Galileo dictionary for direct dark matter detection}},\ }\href
  {https://doi.org/10.1103/PhysRevD.98.123003} {\bibfield  {journal} {\bibinfo
  {journal} {Phys. Rev. D}\ }\textbf {\bibinfo {volume} {98}},\ \bibinfo
  {pages} {123003} (\bibinfo {year} {2018})},\ \Eprint
  {https://arxiv.org/abs/1806.01291} {arXiv:1806.01291 [hep-ph]} \BibitemShut
  {NoStop}%
\bibitem [{\citenamefont {Cirelli}\ \emph {et~al.}(2024)\citenamefont
  {Cirelli}, \citenamefont {Strumia},\ and\ \citenamefont
  {Zupan}}]{Cirelli:2024ssz}%
  \BibitemOpen
  \bibfield  {author} {\bibinfo {author} {\bibfnamefont {M.}~\bibnamefont
  {Cirelli}}, \bibinfo {author} {\bibfnamefont {A.}~\bibnamefont {Strumia}},\
  and\ \bibinfo {author} {\bibfnamefont {J.}~\bibnamefont {Zupan}},\ }\bibfield
   {title} {\bibinfo {title} {{Dark Matter}}\ }\href
  {https://doi.org/10.21468/SciPostPhysRev.1} {10.21468/SciPostPhysRev.1}
  (\bibinfo {year} {2024}),\ \Eprint {https://arxiv.org/abs/2406.01705}
  {arXiv:2406.01705 [hep-ph]} \BibitemShut {NoStop}%
\bibitem [{\citenamefont {Della~Torre}\ \emph {et~al.}(2016)\citenamefont
  {Della~Torre} \emph {et~al.}}]{DellaTorre:2016jjf}%
  \BibitemOpen
  \bibfield  {author} {\bibinfo {author} {\bibfnamefont {S.}~\bibnamefont
  {Della~Torre}} \emph {et~al.},\ }\bibfield  {title} {\bibinfo {title} {{From
  Observations near the Earth to the Local Interstellar Spectra}},\ }in\
  \href@noop {} {\emph {\bibinfo {booktitle} {{25th European Cosmic Ray
  Symposium}}}}\ (\bibinfo {year} {2016})\ \Eprint
  {https://arxiv.org/abs/1701.02363} {arXiv:1701.02363 [astro-ph.HE]}
  \BibitemShut {NoStop}%
\bibitem [{\citenamefont {Boschini}\ \emph {et~al.}(2017)\citenamefont
  {Boschini} \emph {et~al.}}]{Boschini:2017fxq}%
  \BibitemOpen
  \bibfield  {author} {\bibinfo {author} {\bibfnamefont {M.~J.}\ \bibnamefont
  {Boschini}} \emph {et~al.},\ }\bibfield  {title} {\bibinfo {title} {{Solution
  of heliospheric propagation: unveiling the local interstellar spectra of
  cosmic ray species}},\ }\href {https://doi.org/10.3847/1538-4357/aa6e4f}
  {\bibfield  {journal} {\bibinfo  {journal} {Astrophys. J.}\ }\textbf
  {\bibinfo {volume} {840}},\ \bibinfo {pages} {115} (\bibinfo {year}
  {2017})},\ \Eprint {https://arxiv.org/abs/1704.06337} {arXiv:1704.06337
  [astro-ph.HE]} \BibitemShut {NoStop}%
\bibitem [{\citenamefont {Catena}\ and\ \citenamefont
  {Schwabe}(2015)}]{Catena:2015uha}%
  \BibitemOpen
  \bibfield  {author} {\bibinfo {author} {\bibfnamefont {R.}~\bibnamefont
  {Catena}}\ and\ \bibinfo {author} {\bibfnamefont {B.}~\bibnamefont
  {Schwabe}},\ }\bibfield  {title} {\bibinfo {title} {{Form factors for dark
  matter capture by the Sun in effective theories}},\ }\href
  {https://doi.org/10.1088/1475-7516/2015/04/042} {\bibfield  {journal}
  {\bibinfo  {journal} {JCAP}\ }\textbf {\bibinfo {volume} {04}},\ \bibinfo
  {pages} {042}},\ \Eprint {https://arxiv.org/abs/1501.03729} {arXiv:1501.03729
  [hep-ph]} \BibitemShut {NoStop}%
\end{thebibliography}%


\end{document}